\documentclass[epj]{svjour}
\usepackage{graphicx}
\usepackage{amsmath,amssymb}

\begin{document}
\title{Probabilistic Description of Traffic Breakdowns Caused by On-ramp Flow}
\author{Reinhart K\"{u}hne\inst{1} \and
        Ihor Lubashevsky\inst{1,2} \and
        Reinhard Mahnke\inst{3}    \and
        Jevgenijs Kaupu\v{z}s
}                     
\institute{German Aerospace Center, Institute of Transport Research,
           Rutherfordstra{\ss}e 2, 12489 Berlin, Germany, \email{reinhart.kuehne@dlr.de}
      \and
           Theory Department, General Physics Institute, Russian Academy of Sciences,
           Vavilov str., 38, Moscow, 119991, Russia, \email{ialub@fpl.gpi.ru}
      \and
           Fachbereich Physik, Universit\"{a}t Rostock, D--18051 Rostock,
           Germany, \email{reinhard.mahnke@physik.uni-rostock.de}
      \and
          Institute of Mathematics and Computer Science, University of
          Latvia, 29 Rainja Boulevard, LV--1459 Riga, Latvia
}
\date{Received: date / Revised version: date}
%
\abstract{
The characteristic features of traffic breakdown near on-ramp are analyzed. To
describe this phenomenon the probabilistic description regarding the jam
emergence as the formation of a large car cluster on highway inside the
synchronized traffic is constructed. In these terms the breakdown occurs
through the formation of a certain critical nucleus in the metastable vehicle
flow, which is located near the on-ramp. The strong cooperative car interaction
in the synchronized traffic enables us to treat the size of critical jam nuclei
as a large value and to apply to an effective one-lane model. This model
assumes the following. First, the growth of a car cluster is governed by the
attachment of cars to the cluster whose rate is mainly specified by the total
traffic flow. Second, the cluster dissolution is determined by the car escape
from the cluster whose rate depends on the cluster size directly. Third, the
generation of one-car clusters (preclusters) is caused by cars entering the
main road from the on-ramp. The appropriate master equation for the car cluster
evolution is written and the generation rate of critical jam nuclei is found.
The obtained results are in agreement with the empirical facts that the
characteristic time scale of the breakdown phenomenon is about or greater than
one minute and the traffic flow rate interval inside which traffic breakdowns
are observed is sufficiently wide. Besides, as a new results, it is shown that
the traffic breakdown probability can be analyzed, at least approximately,
based solely on the data of the total vehicle flow without separating it into
the vehicle streams on the main road and on-ramp when the relative on-ramp flow
volume exceeds 10\%--20\%.
} 
\maketitle
\section{Introduction}\label{sec:1}

For the last decade physics of traffic flow held attention of physical society
due to two reasons. The former is its obvious importance for traffic
engineering especially concerning the feasibility of attaining the limit
capacities of traffic networks and quantifying it. The latter is related to the
fact that vehicle ensembles on highways form a sufficiently simple example of
systems with motivation being the object of new branches in modern physics.
Indeed, on one hand, the individual motion of cars is affected essentially by
the driver behavior in addition to the regularities of classical mechanics. So,
in this sense, the vehicle ensembles are nonphysical systems. On the other
hand, on the macroscopic level the vehicle ensembles exhibit a lot of
properties like phase formation and phase transitions widely met in physical
systems (for a review see \cite{rev1,rev2,Helbing}).

The traffic breakdown, i.e. the initial stage of jam formation typically near
bottlenecks is an important phenomenon for traffic engineering, exactly its
main characteristics determine the limit capacity of the corresponding road
fragments or nodes. Its properties are sufficiently complex, traffic breakdown
usually proceeds through the sequence of two phase transitions: free flow $\to$
synchronized traffic $\to$ stop-and-go pattern with a number of hysteresis and
nucleation effects \cite{K2,KB2}. A detailed description of the jam formation
near bottlenecks can be found in Ref~\cite{K3}. Nevertheless the basic
properties of traffic breakdown are far from being understood well. In
particular, it is a probabilistic phenomenon \cite{LilyIn,Persaud}, i.e.
traffic breakdown occurs not immediately after the vehicle flow rate attaining
a certain critical value but randomly within some interval $(q_{c1},q_{c2})$.
It is easy to explain this assuming homogeneous traffic flow to be metastable
and a jam to emerge via the nucleation mechanism. A nontrivial fact is a large
width of the traffic breakdown interval. According to the empirical data
\cite{Persaud,Lily} $q_{c1} \sim 1000$--1500~v/h/l (vehicle per hour per lane)
and $q_{c2} \sim 1900$--2800~v/h/l, i.e. the ratio $(q_{c2} -q_{c1})/q_{c1}\sim
50$\%--100\%. The traffic breakdown phenomenon is also characterized by a large
time interval during which it develops. Namely, it is about several minutes, a
typical observation time of detecting traffic breakdown according to the
traffic engineering technique. By contrast, the characteristic time scale of
individual car dynamics is about or less than ten seconds.

\begin{figure}
\begin{center}
\includegraphics[scale=1]{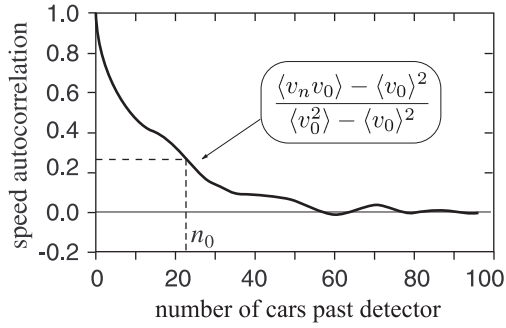}
\end{center}
\caption{Illustration of the speed autocorrelation \textsl{vs} the number of
        cars that have passed a fixed detector. Based on the observations by
        Neubert, Santen, Schadschneider, and Schreckenberg \cite{Newman}.}
 \label{FI}
\end{figure}

In the previous paper \cite{we1} we have developed a simple model explaining
the two last features of traffic breakdown. Its basic point is the assumption
that the synchronized phase of traffic flow develops first and traffic
breakdown, in its own accord, comes into being via the formation of a critical
jam nucleus inside this phase with strong cooperative interaction between cars.
This corresponds to the recent notion about the characteristic properties of
traffic flow near bottlenecks \cite{K2,KB2,K3}. The size $n_0$ of such critical
nuclei must be sufficiently large, $n_0\sim 10$--20, which is justified by the
analysis of single vehicle data \cite{Newman} and illustrated in Fig.~\ref{FI}.
As seen in Fig.~\ref{FI} a car cluster in the synchronized traffic must span
over many cars along the lane. Such a car cluster also can span over all the
lane on a highway, which enable us to apply to an effective one-lane
approximation dealing with macrovehicles rather than real individual cars.

This model, however, has left the question about the jam origin or, what is the
same, about the source of one-car clusters (preclusters) beyond the
consideration. Typically traffic breakdowns are caused by the influence of
different bottlenecks, in particular, on- or off-ramps. Exactly vehicles
entering or leaving traffic flow on the main road are source of jam nuclei.
Investigation of the effect of the on-ramp flow on the traffic breakdown is the
subject of the present work. As a particular result we would like to find an
justification for the analysis of traffic breakdown phenomena near on-ramps
using only the data of the total traffic flow leaving the on-ramp region. This
is a typical way of analyzing real empirical data because of the lack of
statistics.

\section{Model: Car cluster growth near on-ramp}\label{sec:2}

\begin{figure}
\begin{center}
\includegraphics[width=80mm]{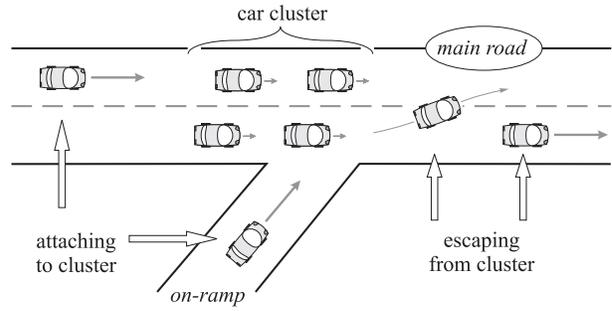}
\end{center}
 \caption{The mechanism under consideration that governs the formation and
  growth of a car cluster on a highway near an on-ramp}
 \label{F1}
 \end{figure}

The proposed model for the formation of a car cluster caused by on-ramp vehicle
flow and its further growth is illustrated in Fig.~\ref{F1}. The model is based
on the following assumptions. First, each car entering the highway from the
on-ramp either attaches itself to an existing car cluster or forms a precluster
(one-car cluster) which further can grow or dissipate. Second, when a car
moving on the highway reaches the cluster it attaches itself to the cluster.
Third, due to the lane change maneuvers the escape rate of cars from the
cluster depends on the cluster size. Fourth, the formed car cluster is
characterized by a small mean velocity so it is practically localized near the
on-ramp.

The given model takes into account that a real jam nucleus develops inside the
synchronized traffic on a multilane highway and it exhibits irreversible growth
only after its size $n$ attaining a sufficiently large critical value
$n_c\gg1$. Due to the strong multilane cooperative effects in the synchronized
traffic all the vehicles interact with suhc a cluster. For a small jam nucleus
real cars either can avoid it by changing the lanes or leave the cluster fast
because its mean velocity has no time to drop sufficiently low. This is
formally allowed for by the dependence of the escape rate on the cluster size
$n$. So the proposed effective one-lane model imitates a more complex
phenomenon of traffic breakdown on highways.

\begin{figure}
\begin{center}
\includegraphics[width=65mm]{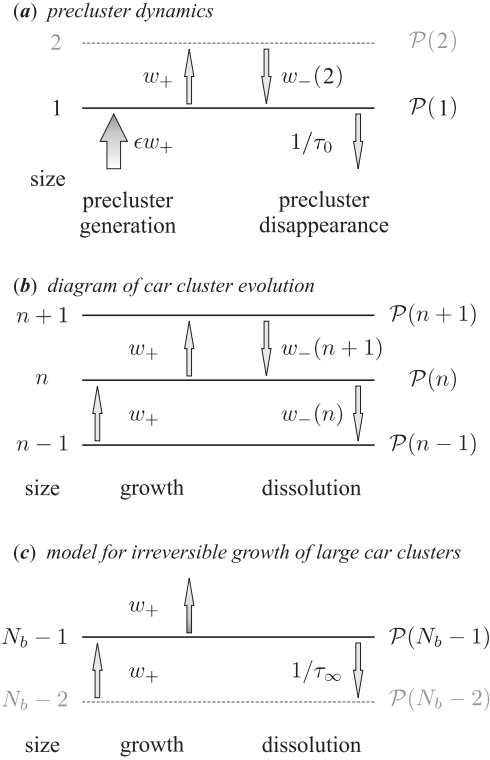}
\end{center}
\caption{Schematic illustration of the cluster transformations.}
\label{F2}
\end{figure}

\begin{figure}
\begin{center}
\includegraphics[width = 65mm]{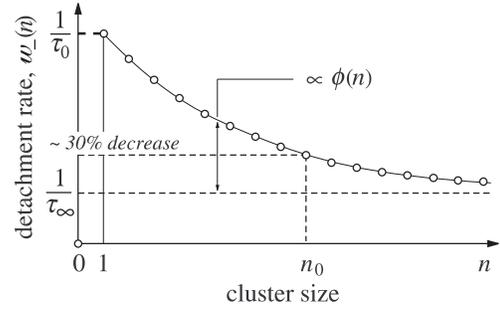}
\end{center}
\caption{The detachment rate $w_{-}(n)$ \textsl{vs} the cluster size $n$. A
qualitative sketch.} \label{F3}
\end{figure}

The adopted assumptions lead to the scheme of the car cluster dynamics shown in
Fig.~\ref{F2}. Each car attachment to or detachment from the cluster causes
unit change of its size whose sequence is considered to be mutually
independent. The attachment rate to the cluster of size $n\geq1$ is written in
the form
\begin{equation}
  \label{2.1}
    w_+  = q_\text{in} + q_\text{ramp}\,,\\
\end{equation}
where $q_\text{in}$ is the rate of vehicle flow on the main road entering the
on-ramp region (per lane) and $q_\text{ramp}$ is the rate of traffic flow on
the on-ramp. The rate of one-cluster generation (precluster generation) is
\begin{equation}\label{2.3}
    w_+^0 = q_\text{ramp} := \epsilon w_+\,.
\end{equation}
Here the latter equality is no more than the definition of the value $\epsilon$
treated as a small parameter, $\epsilon\ll1$. Expression~(\ref{2.3}) is the
mathematical implementation of the assumptions that the car clusters are
initially generated solely by cars from the on-ramp and, thus, without the
on-ramp vehicle flow no traffic breakdown is possible. Finally, the detachment
rate from the car cluster of size $n$ is specified by the \textit{Ansatz}
\begin{equation}\label{2.4}
  w_{-}(n)=\frac{1}{\tau (n)}:=\frac{1-\phi (n)}{\tau _{\infty }}+
  \frac{\phi(n)}{\tau _{0}}\,,
\end{equation}
where the function $\phi(n)$ decreases from 1 to 0 as the cluster size $n$ runs
from 1 to $\infty$. Time scales $\tau_0 < \tau_\infty$ characterize the
detachment rate from small clusters and sufficiently large ones, respectively.
Their values can be estimated from the relations $q_{c1} \approx 1/\tau_\infty$
and $q_{c2} \approx 1/\tau_0$, where $q_{c1}$ and $q_{c2}$ are the traffic flow
rates (per lane) such that a jam cannot form at all for $q< q_{c1}$ and a jam
emerges immediately when the flow rate $q$ exceeds $q_{c2}$ (see
Section~\ref{sec:1}). In these estimates we do not distinguish between the flow
rate on the main road $q_{\text{in}}$ and the total flow rate
$q_\text{in}+q_{\text{ramp}}$ because typically $q_{\text{ramp}}\ll
q_\text{in}$. Applying to the available empirical data \cite{Persaud,Lily} we
set, for example, $q_{c1}\sim 1500$~v/h/l and $q_{c2}\sim 2500$~v/h/l whence
get $\tau_0\sim 1.5$~sec and $\tau_\infty\sim 2.5$~sec. Besides, as also
discussed in Section~\ref{sec:1} a scale $n_0$ dividing the car clusters into
small and large ones is much greater than unity, $n_0\sim 10$--20, due to the
cooperative interaction of cars in the synchronized traffic. The form of the
$w_{-}(n)$-dependence is schematically shown in Fig.~\ref{F3}.

To be specific in obtaining some numerical estimates the following simple
\textit{Ansatz}
\begin{equation}\label{2.4phi}
    \phi(n) = \frac{n_0+1}{n_0+n}\,,
\end{equation}
will be used, however, the particular form of the $\phi(n)$-dependence is of
minor effect.

The car cluster evolution is described by the dynamics of the distribution
function $\mathcal{P}(n,t)$ which, by virtue of Diagram~\ref{F2}, obeys the
forward master equation (following Mahnke \emph{et al.} \cite{M1,M2})
\begin{eqnarray}
 \nonumber
 \partial_t \mathcal{P}(n,t) & = & w_+ \mathcal{P}(n-1,t) +
 w_{-}(n+1)\mathcal{P}(n+1,t)\\
 & &{} - [w_+ + w_{-}(n)] \mathcal{P}(n,t)
\label{2.5}
\end{eqnarray}
for $n\geq2$. The precluster evolution is described by the equation
\begin{equation}\label{2.6}
 \partial_t \mathcal{P}(1,t) = w_+^0 - [w_+ + w_{-}(1)] \mathcal{P}(1,t)\,.
\end{equation}
The model under consideration describes only the initial stage of jam
emergence, i.e. the formation of the jam critical nucleus. When the size $n$ of
a car cluster exceeds a certain critical value $n_c$ it undergoes the
irreversible growth giving rise to the jam formation. Within the frameworks of
the adopted description this effect is taken into account by the following
``boundary'' condition imposed on the distribution function $\mathcal{P}(n,t)$
taken at a sufficiently distant point $N_b\gg n_0$:
\begin{equation}\label{2.7}
    \mathcal{P}(N_b,t) = 0\,.
\end{equation}
Naturally, in this case equation~(\ref{2.5}) holds at points $2\leq n \leq N_b
-1$. The system of equations~(\ref{2.5})--(\ref{2.7}) makes up the proposed
model.

In what follows the assumption
\begin{equation}\label{2.eq}
  \frac{1}{\tau_\infty} \approx q_{c1}  < q_{\text{in}} < q_{c2} \approx \frac{1}{\tau_0}
\end{equation}
will be adopted. In other words, we will confine our consideration to the case
when the traffic breakdown can occur but the homogeneous state of the vehicle
flow is locally stable. Exactly in this case the traffic breakdown exhibits the
probabilistic behavior. Then assuming the traffic flow rate on the on-ramp and
the main road to be fixed the given model enables us to calculate directly the
generation rate of jam critical nuclei $G_c$, what is done in
Appendix~\ref{app}. Using the obtained formula~(\ref{app.9}) and the
relation~(\ref{2.3}) the desired value $G_c$ can be written as
\begin{equation}\label{2.8}
    G_c \approx q_\text{ramp}\sqrt{\frac{\beta_c}{2\pi n_c}}
 \exp\left\{-\Omega(n_c)\right\}\,.
\end{equation}
Here the critical car cluster size $n_c$ is specified by
equation~(\ref{app.6}), $\Omega(n_c)$ is the ``growth''
potential~(\ref{app.6Omega}) taken at the point $n=n_c$, and the constant
$\beta_c\sim 1$ is determined by expansion~(\ref{app.7}). The car clusters have
to overcome exactly the potential barrier $\Omega(n_c)$ for their growth to
become irreversible.

\begin{figure}
\begin{center}
\includegraphics[width = 70mm]{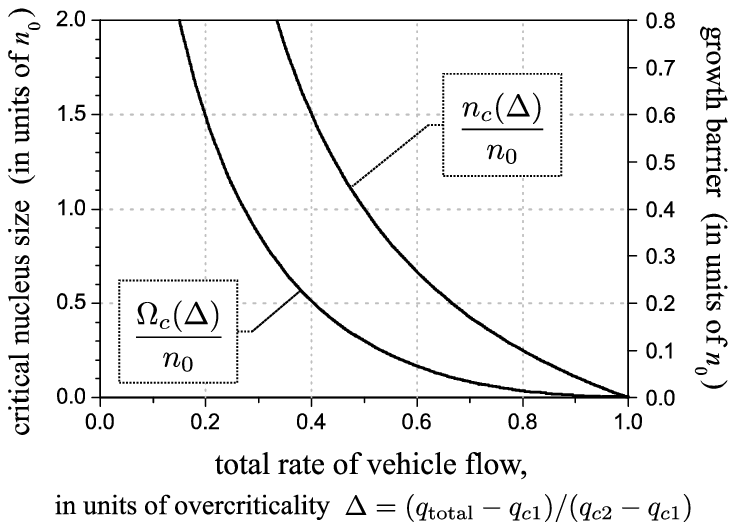}
\end{center}
\caption{The size of critical jam nuclei, $n_c$, and the growth barrier
$\Omega_c$ vs the total traffic flow rate $q_\text{total} = q_\text{in} +
q_\text{ramp}$ measured in the units of the overcriticality $\Delta =
(q_\text{total}-q_{c1})/(q_{c2}-q_{c1})$. In obtaining these curves we used
\emph{Ansatz}~(\ref{2.4phi}) and set $(q_{c2}-q_{c1})/q_{c1} = 1.0$.}
\label{F4}
\end{figure}

\begin{figure}
\begin{center}
\includegraphics[width = 70mm]{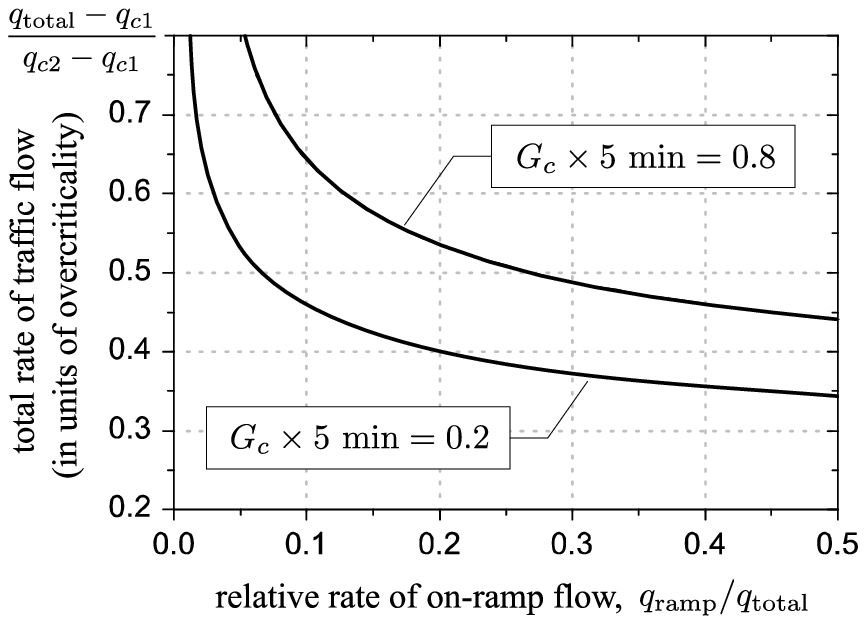}
\end{center}
\caption{Solid curves on the phase plane ``on-ramp flow rate -- total vehicle
flow rate'' depict the loci where the generation rate $G_c$ of jam critical
nuclei takes the values 0.2/5~min and 0.8/5~min, respectively. Roughly
speaking, these curves correspond to points on this phase plane where the
probability of traffic breakdown within 5-min observation interval is about
20\% and 80\%, respectively. In obtaining these curves we used
\emph{Ansatz}~(\ref{2.4phi}) and set $q_{c1} = 1500$~veh/h/l, $q_{c2} =
3000$~veh/h/l, and $n_0 = 10$.} \label{F5}
\end{figure}

Figure~\ref{F4} depicts the dependence of the critical cluster size
$n_c(\Delta)$ as well as the growth barrier $\Omega(n_c) :=\Omega_c(\Delta)$ on
the total traffic flow rate $q_\text{total} := q_\text{in} + q_\text{ramp}$ in
units of the overcriticality
\begin{equation}\label{2.100}
    \Delta := \frac{q_\text{total}-q_{c1}}{q_{c2}-q_{c1}}
\end{equation}
that were obtained using \emph{Ansatz}~(\ref{2.4phi}). The corresponding form
of the critical cluster generation rate $G_c$ is (see Appendix~\ref{app},
expression~\eqref{app.last})
\begin{equation}\label{2.9}
    G_c = \frac{q_\text{ramp}}{\sqrt{2\pi n_0}}
    \sqrt{\frac{q_{c2}-q_{c1}}{q_\text{total}}}
    \left(\frac{q_{c2}}{q_\text{total}}\right)^{\tfrac{n_0q_{c2}}{q_{c1}}}
    \Delta^{1+\tfrac{n_0(q_{c2}-q_{c1})}{q_{c1}}}\,.
\end{equation}
Figure~\ref{F5} exhibits the characteristic features of the obtained critical
cluster generation rate depending on the volumes of the on-ramp flow and the
total vehicle flow.

\section{Conclusion}

The present paper developed a probabilistic description of the traffic
breakdown phenomena near an on-ramp. Previously \cite{we1} we have proposed a
simple probabilistic model for the traffic breakdown that explains two facts
observed empirically (see Refs~\cite{Persaud,Lily} as well for a review
Ref.~\cite{Helbing}). The former is the large width of the traffic flow rate
interval $(q_{c1},q_{c2})$ wherein the traffic breakdown phenomena are
observed, namely, $(q_{c2}-q_{c1})/q_{c1}\sim 50\%$--100\%. The latter is the
fact that the time interval during which traffic breakdown develops is about
several minutes whereas the characteristic time scale of the individual car
dynamics is about or less than ten seconds. However the origin of traffic
breakdown has been left beyond the analysis. Typically different kinds of
bottlenecks, for example, on- and off-ramps, are responsible for the jam
emergence. The main purpose of the given paper was to take into account
directly the effect of on-ramp within this probabilistic description. Besides
we would like to explain why the empirical analysis tackling the traffic
breakdown near on-ramps can be performed dealing solely with the data of the
total traffic flow without separating it into main road and on-ramp streams. It
is a typical situation because of the lack of statistics.

According to the modern notion of jam emergence it proceeds mainly through the
sequence of two phase transitions: free flow $\to$ synchronized mode $\to$
stop-and-go pattern \cite{KB2}. Both of these transitions are of the first
order, i.e. they exhibit breakdown, hysteresis, and nucleation effects
\cite{K2}. However, in the jam formation the second transition typically plays
the leading role (for a detailed analysis see Ref.~\cite{K3}).

Therefore the proposed model assumes, as the basic point, that a jam develops
inside a certain mode of traffic flow with strong cooperative car interaction.
In other words, cars entering the main road from the on-ramp give rise to a
cooperative phase of car motion. Exactly inside this phase jam nuclei occur and
lead to the irreversible jam formation when their size exceeds randomly a
certain critical value. Due to the cooperative car interaction this critical
size must be much larger than unity, as it follows from the available single
vehicle data~\cite{svd}. Besides, in this case a critical jam nucleus has to
span over all the lane, which enabled us to use an effective one-lane
approximation actually dealing with macrovehicles rather than real cars.

The effect of complex interaction between the cars moving on the main road and
entering the neighborhood of the on-ramp with a jam nucleus located near the
on-ramp is taken into accounts as follows. A jam nucleus is treated as a
cluster of car moving sufficiently slow near the on-ramp. Each car entering the
on-ramp neighborhood with a car cluster attaches itself to it. In this approach
the rate of the cluster growth $w_+$ is considered to be determined completely
by the vehicle flow on the main road entering the on-ramp region as well as by
the traffic flow on the on-ramp. The car detachment process is described by the
escaping rate $w_-(n)$ depending on the cluster size $n$ and decreasing with
$n$. Therefore both the facts that cars can avoid a small jam nucleus by
changing the lanes and overtaking it as well as can escape it also changing the
lanes are allowed for by the $w_-(n)$-dependence.

What is new in the model under consideration in comparison with the previous
one \cite{we1} is the assumption that one-car clusters, i.e. car preclusters
being the initial state of jam nuclei are due to cars entering the main road
from the on-ramp. Expression~(\ref{2.8}) or its particular form~(\ref{2.9})
specifies the desired generation rate $G_c$ of the critical jam nuclei
depending, in particular, on the on-ramp flow rate. The obtained result is
illustrated in Fig.~(\ref{F5}).

The main conclusion of the present paper is the following. When the vehicle
flow rate, $q_\text{ramp}$, exceeds 10\%--20\% of the total traffic flow rate,
$q_\text{total}$, the characteristics of the traffic breakdown depend weakly on
$q_\text{ramp}$ individually. So in this case the traffic breakdown phenomenon
can be analyzed, at least semiqualitatively, using solely the total flow rate
data. Otherwise, $q_\text{ramp}\lesssim q_\text{total}$, the details of
partitioning the traffic flow rate into the main and on-ramp streams are the
factor. In addition, the given model, as previous one \cite{we1}, explains the
large width of the vehicle flow rate interval $(q_{c1}, q_{c2})$ wherein
traffic breakdowns are observed. It relates the critical values $q_{c1}$,
$q_{c2}$ to the dependence of the car detachment rate $w_-(n)$ on the cluster
size $n$, namely, $q_{c1} = w_-(\infty)$ and $q_{c2}= w_-(1)$. Since a car
cluster forms inside the synchronized traffic where the car cooperative
interaction is strong the characteristic value $n_0$ separating car clusters
into ``small'' and ``large'' is much greater than unity. Therefore the ratio
$(w_{1}-w_{\infty})/w_{\infty}$ has to be about unity. Besides, as follows from
expression~(\ref{2.8}) the characteristic time scale of the traffic breakdown
development is about $\tau_{\text{bd}}\sim \sqrt{2\pi
n_0}(q_{c1}/q_{\text{ramp}})\tau_\infty\gtrsim 1$~min for $q_{\text{ramp}}\sim
0.2q_{c1}$. This estimate explains ones more why the traffic breakdown
phenomena are typical detected within 5--15 minute intervals.

\begin{acknowledgement}

This work was supported in part by RFBR Grants~02-02-16537, Russian Program
``Integration'' Project B0056, and DFG Grant MA 1508/6.

\end{acknowledgement}

\appendix
\section{Steady-state flux of jam nuclei}\label{app}

\begin{figure}
\begin{center}
\includegraphics[scale=1]{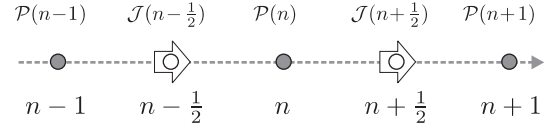}
\end{center}
\caption{Illustration of the car cluster ``space'' and the quantities defined
 in it. The solid circles depict the integers $\{n\}$, the cluster size
 values, at which the distribution function $\mathcal{P}(n,t)$ is defined. The
 hole circles match the intermediate points $\{n+\frac12\}$ at which the cluster
 flux $\mathcal{J}(n+\frac12\,,t)$ is considered.}
 \label{Fapp1}
\end{figure}

The system of equations~(\ref{2.5})--(\ref{2.7}) describes the evolution of jam
nuclei at the initial stage of jam emergence. In terms of the cluster flux
$\mathcal{J}(n+\frac12\,,t)$ defined in the car cluster ``space''
(Fig.~\ref{Fapp1})
\begin{equation}\label{app.1}
   \mathcal{J}(n+{\tfrac12}\,,t) := w_+ \mathcal{P}(n,t) - w_-(n+1) \mathcal{P}(n+1,t)
\end{equation}
this model can be reduced to the governing equation holding for $1\leq n \leq
N_b-1$
\begin{equation}\label{app.2a}
    \partial_t\mathcal{P}(n,t) = \mathcal{J}(n-{\tfrac12}\,,t) -
    \mathcal{J}(n+{\tfrac12}\,,t)
\end{equation}
subject to the following two ``boundary'' conditions
\begin{eqnarray}
  \label{app.2b}
  \mathcal{J}({\tfrac12}\,,t) &=& \epsilon w_+ - w_-(1)\mathcal{P}(1,t)\,, \\
  \label{app.2c}
  \mathcal{J}(N_b-{\tfrac12}\,,t) &=& w_+\mathcal{P}(N_b-1,t)\,.
\end{eqnarray}
Here equation~(\ref{app.2a}) is no more than equation~(\ref{2.5}) rewritten in
the new terms, whereas values~(\ref{app.2b}) and (\ref{app.2c}) are the
precluster flux and the flux of large car cluster caused by their irreversible
growth, respectively. The two expressions stem directly from
equations~(\ref{2.6}), (\ref{2.7}) and are illustrated in
Fig.~\ref{F2}\emph{a,c}.

The steady-state solution of system~(\ref{app.2a})--(\ref{app.2c}), in
particular, the steady-state value of the cluster flux $G_c :=
\mathcal{J}(N_b-\frac12)$ gives us the desired generation rate of jam critical
nuclei. Naturally, in this case the on-ramp flow rate as well as the traffic
flow rate on the main road have to be treated as constant in time values.

The steady-state condition leads to a constant value of the car cluster flux
$\mathcal{J}(n-\frac12) = G_c$ which together with the ``boundary''
condition~(\ref{app.2c}) gives us the relation
\begin{equation}\label{app.3}
   \mathcal{P}_s(n)  =  \left\{1 + \sum_{p=1}^{N_b-1-n}\prod_{q=1}^{p}
   \frac{w_-(n+q)}{w_+} \right\}\frac{G_c}{w_+}\,.
\end{equation}
The substitution of (\ref{app.3}) into (\ref{app.2b}) yields the desired
expression for the generation rate of jam critical nuclei
\begin{equation}\label{app.4}
    G_c = \epsilon w_+\left\{1+ \sum_{p=1}^{N_b-1}\exp\left[ \sum_{q=1}^{p}
    \ln\left(\frac{w_-(q)}{w_+}\right)\right] \right\}^{-1},
\end{equation}
which is the main result of Appendix~\ref{app}.

\subsection*{Continuum approximation}

\begin{figure}
\begin{center}
\includegraphics[width = 70mm]{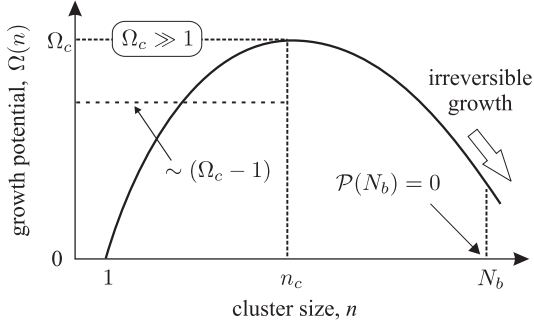}
\end{center}
\caption{Schematic illustration of the growth potential and the corresponding
characteristic values of the cluster size.}
 \label{Fapp2}
\end{figure}

When the inequalities
\begin{gather}
  \label{app.5a}
    w_-(\infty) < w_+ < w_-(1)\,,
  \\ 
  \label{app.5b}
    (w_+ - w_-(\infty))n_0\,,\, (w_-(1)-w_+ )n_0 \gg 1
\end{gather}
hold expression~(\ref{app.4}) is simplified. Since the characteristic scale
$n_0$ on that the quantity $w_-(n)$ exhibits substantial decrease is large,
$n_0\gg 1$, the sums in expression~(\ref{app.4}) can be replaced by the
corresponding integrals. Indeed, let us introduce the ``growth potential''
\begin{equation}\label{app.6Omega}
    \Omega(n) := \sum_{q=1}^n \ln\left(\frac{w_-(q)}{w_+}\right).
\end{equation}
Its form is illustrated in Fig.~\ref{Fapp2}. In particular, the growth
potential attains the maximum at the point $n_c$ being the root of the equation
\begin{equation}\label{app.6}
    w_-(n_c) = w_+
\end{equation}
and playing the role of the critical size of jam nuclei.

In the vicinity of the point $n_c$ the value $w_-(n)$ can be treated as a
function of the continuous argument $n$ and approximated by the expression
\begin{equation}\label{app.7}
    w_-(n) \approx w_+\left[1 - \beta_c \frac{(n-n_c)}{n_c}\right],
\end{equation}
where $\beta_c\sim 1$ is a constant about unity. Since the main contribution to
the first sum in expression~(\ref{app.4}) is due to a certain neighborhood of
the point $n_c$ we approximate the growth potential as follows
\begin{equation}\label{app.8}
  \Omega(n) = \Omega(n_c) - \beta_c\frac{(n-n_c)^2}{2n_c}\,.
\end{equation}
Substituting formula~(\ref{app.8}) into expression~(\ref{app.4}) and replacing
the sum running over $p$ by the corresponding integral over the continuous
variable $p$ we get
\begin{equation}\label{app.9}
 G_c \approx \epsilon w_+ \sqrt{\frac{\beta_c}{2\pi n_c}}
 \exp\left\{-\Omega(n_c)\right\}\,.
\end{equation}
Formula~(\ref{app.9}) is the desired expression for the generation rate of jam
critical nuclei.

In particular, in the given limit for \textit{Ansatz}~\eqref{2.4} the critical
size $n_c$ of jam nuclei is specified by the expression
\begin{equation}\label{app.add1}
    \phi(n_c) = \Delta\,,
\end{equation}
where the traffic flow overcriticality measure $\Delta$ was introduced by
formula~\eqref{2.100} and we have set $q_{c1} = 1/\tau_\infty$ and $q_{c2} =
1/\tau_0$. Using in addition \textit{Ansatz}~\eqref{2.4phi} we get from
formulae~\eqref{app.6Omega} and \eqref{app.7} the expressions for the parameter
\begin{gather}
    \label{app.add2}
    \frac{\beta_c}{n_c}  =
    \frac{(q_{c2}-q_{c1})}{q_\text{total}}\,\frac{\Delta^2}{n_0}
    \\
\intertext{and for the critical potential barrier}
    \label{app.add3}
    \Omega(n_c) = -n_0\left[\frac{(q_{c2}-q_{c1})}{q_{c1}}\ln \Delta
    + \frac{q_{c2}}{q_{c1}}\ln\Big(\frac{q_{c2}}{q_\text{total}}\Big)
    \right]\,.
\end{gather}
Substitution of expressions~\eqref{app.add2} and \eqref{app.add3} into
\eqref{app.9} yields
\begin{equation}\label{app.last}
    G_c \approx \frac{\epsilon w_+}{\sqrt{2\pi n_0}}
    \sqrt{\frac{q_{c2}-q_{c1}}{q_\text{total}}}
    \left(\frac{q_{c2}}{q_\text{total}}\right)^{\tfrac{n_0q_{c2}}{q_{c1}}}
    \Delta^{1+\tfrac{n_0(q_{c2}-q_{c1})}{q_{c1}}}\,,
\end{equation}
which is exactly formula~\eqref{2.9} due to relation~\eqref{2.3}.

\end{document}